\documentclass[12pt,a4paper]{article}
\usepackage{amsmath,epsfig}
\begin{document}
 
\begin{titlepage}
\renewcommand{\thefootnote}{\fnsymbol{footnote}}
\begin{flushright}
MPI-PhT/2000-44\\
\end{flushright}

\vspace{1cm}

\begin{center}
{\LARGE Kosterlitz-Thouless theory and lattice artifacts}
\vspace{1.6cm}
 
{{\large J. Balog\footnote{on leave of absence from the Research 
Institute for Particle and Nuclear Physics, Budapest, Hungary}\\[8mm]
{\small\sl Max-Planck-Institut f\"{u}r Physik}\\
{(\small\sl Werner Heisenberg Institut)}\\[3mm]
{\small\sl F\"{o}hringer Ring 6, 80805 Munich, Germany}}}

\vspace{2.4cm}
{\bf Abstract}
\end{center}

\begin{quote}
The massive continuum limit of the $1+1$ dimensional $O(2)$
nonlinear $\sigma$-model (XY model) is studied using its
equivalence to the Sine-Gordon model at its asymptotically free
point. It is shown that leading lattice artifacts are universal but
they vanish only as inverse powers of the logarithm of the correlation 
length. Such leading artifacts are calculated for the case of the scattering
phase shifts and the correlation function of the Noether current using
the bootstrap S-matrix and perturbation theory respectively.
\end{quote}
\vfill
\setcounter{footnote}{0}
\end{titlepage}

\newcommand{\be}{\begin{equation}}
\newcommand{\ba}{\begin{eqnarray}}
\newcommand{\ea}{\end{eqnarray}}

\section{Introduction}
In this paper we study the properties of the two-dimensional
$O(2)$ nonlinear $\sigma$-model, better known as the XY model.
This model has been the subject of extensive theoretical and numerical
analysis, starting with the seminal papers of Kosterlitz and Thouless (KT)
\cite{KT}. For a review of KT theory, see \cite{ZinnJ}.

Analitical work is usually based on a series of mappings that starts at
the original (lattice) XY model and arrives at the Sine-Gordon (SG) model
or its fermionic equivalent, a (deformed) version of the chiral Gross-Neveu
(CGN) model. This latter formulation is most useful if one wants to study
questions related to the dynamically generated $SU(2)$ symmetry of the model.

Most papers on the XY model study its properties interesting for
Statistical Physics, in particular the pecularities of the KT phase
transition, which is of infinite order. In this paper we look at the XY
model as an example of $1+1$ dimensional relativistic Quantum Field Theory. 
More precisely, we study the massive continuum limit of the lattice theory,
which, in the language of Statistical Physics, means that we approach the
KT phase transition point from the high temperature phase.

Treating the XY model as the $n=2$ member of the family of $O(n)$
nonlinear $\sigma$-models gives additional insights, since a lot is
known about the $n\geq3$ models~\cite{ZZ}. More importantly, using the
SG language, we show that the approach to the continuum limit in this
model is much slower than in most other lattice models. Lattice artifacts
vanish in this model, instead of the
usual Symanzik type behaviour~\cite{Sym} (i.e. integer powers of the 
lattice spacing), as inverse powers of the logarithm of the lattice 
spacing only. On the other hand, we can show that the leading artifacts
are universal and calculable. Our main result is Eqs. (\ref{gammaexp})
and (\ref{xiexp}) in Section 4, which allow us to calculate leading
lattice artifacts in terms of SG data.

We can make use of the fact that the SG model is exactly solvable
and its bootstrap S-matrix is exactly known. We calculate the leading 
artifacts for the scattering phase shifts using the bootstrap results.
An alternative method is perturbation theory (PT). Since the SG model
is asymptotically free if we use suitable expansion parameters
\cite{Amit}, the methods of renormalization group (RG) improved PT
are thus available.

In Section 2 we review the relation of the XY model to the $O(n)$ models
with $n\geq3$ and describe the chain of mappings leading from the XY model
to the SG model and its equivalent fermionic formulation.

In Section 3 we recall the analysis of the phase diagram of the model
in the vicinity of the KT phase transition point. This is described in the
SG language.

In Section 4 we explain how to calculate the lattice artifacts and apply
this to the case of the scattering phase shifts.

Finally in Section 5 we calculate the lattice artifacts for the 
two-point correlation function of the Noether current corresponding
to the $O(2)$ symmetry. Here we use the method of RG improved PT.
To calculate the value of a non-perturbative constant needed here,
we also consider the system in the presence of an external field coupled
to the Noether charge. (This calculation is analogous to the one used
previously to determine the $M/\Lambda$ ratio for the $O(n)$ models
\cite{HMN}.) We give here a parameter-free two-loop formula for the
lattice artifacts. 

A precision MC study of the massive continuum limit of the $O(2)$ model
will be described in a forthcoming paper \cite{paper3}.

\section{From the XY model to the Sine-Gordon model and beyond}

In this section we describe in some detail the chain of mappings starting
with the XY model and ending at the SG model and its fermionic equivalent.

We can treat the XY model as the $n=2$ member of the family of
$O(n)$ nonlinear $\sigma$-models with Lagrangean
\be
{\cal L}^{O(n)}=\frac{1}{2g^2}\,\partial_\mu S^a\partial_\mu S^a
\quad ;\quad S^aS^a=1,\qquad a=1,2,\dots n.
\label{LagOn}
\end{equation}
The $n\geq3$ models are known to be integrable. Polyakov \cite{Poly}
and L\"uscher \cite{Lusch} have shown the existence of 
respectively local and nonlocal higher spin conserved charges, whose
existence implies quantum integrability. Assuming the spectrum of the
model consisted of an $O(n)$ vector multiplet of massive particles the
exact S-matrix of the $n\geq3$ models was found by bootstrap methods
\cite{ZZ}:
\be
S_{ab;cd}(\theta)=
\sigma_1(\theta)\delta_{ab}\delta_{cd}+
\sigma_2(\theta)\delta_{ac}\delta_{bd}+
\sigma_3(\theta)\delta_{ad}\delta_{bc},
\label{OnSmatrix}
\end{equation}
where
\begin{alignat}{1}
\sigma_1(\theta)&=\frac{-2\pi i\theta}{i\pi-\theta}
\,\cdot\,\frac{s^{(2)}(\theta)}
{(n-2)\theta-2\pi i}\,,\nonumber\\
\sigma_2(\theta)&=(n-2)\theta
\,\cdot\,\frac{s^{(2)}(\theta)}
{(n-2)\theta-2\pi i}\,,\label{sig2}\\
\sigma_3(\theta)&=-2\pi i
\,\cdot\,\frac{s^{(2)}(\theta)}
{(n-2)\theta-2\pi i}\nonumber
\end{alignat}
and the \lq isospin 2' phase shift $s^{(2)}$ is given by
\be
s^{(2)}(\theta)=-\exp\left\{2i\int_0^\infty\frac{d\omega}{\omega}
\sin\omega\theta \tilde K_n(\omega)\right\}
\label{iso2}
\end{equation}
with
\be
\tilde K_n(\omega)=
\left[\frac{e^{-\pi\omega}+e^{-2\pi\frac{\omega}{n-2}}}
{1+e^{-\pi\omega}}\right].
\label{Kn}
\end{equation}

Much less is known about the $O(2)$ model. A simple observation
is that (\ref{sig2}) and also (\ref{Kn}) have a smooth
$n\to2$ limit. It is natural to assume that the $O(2)$ model is
also integrable, its spectrum consists of a single $O(2)$ doublet
of massive particles whose scattering is indeed described by the
$n\to2$ limit of the S-matrix (\ref{OnSmatrix}).

Although taking the formal $n\to2$ limit of the bootstrap results
valid for $n\geq3$ is not convincing in itself, the conclusion
turns out to be correct because as we will see it also follows from 
the Kosterlitz-Thouless theory \cite{KT} of the XY model. 

Before turning to the KT theory we make a small digression to
discuss the two-dimensional Sine-Gordon (SG) model.
Its Lagrangean can be written as
\begin{equation}
{\cal L}^{\rm SG}=\frac12\partial_\mu\phi\partial_\mu\phi
+\frac{\alpha}{\beta^2a^2}\left[1-\cos(\beta\phi)\right],
\label{LagSG}
\end{equation}
where $\alpha$ is the dimensionless mass parameter, $a$ is
a constant of dimension mass$^{-1}$ and $\beta$ is
the SG coupling. It is also integrable and its spectrum and S-matrix
was also found in \cite{ZZ}. The spectrum depends on $\beta$ 
in a complicated way but it becomes simple for the range 
$8\pi>\beta^2>4\pi$ when it is free of any bound states
and consists of a single $O(2)$ vector of massive particles whose S-matrix
can again be written as (\ref{OnSmatrix}) but now
\begin{alignat}{1}
\sigma_1(\theta)&=\cosh\frac{i\pi\nu}{2}
\sinh\frac{\theta\nu}{2}\,\,\frac{s^{(2)}_{\rm SG}(\theta)}
{\sinh\frac{(i\pi-\theta)\nu}{2}}\,,\nonumber\\
\sigma_2(\theta)&=-\sinh\frac{i\pi\nu}{2}
\sinh\frac{\theta\nu}{2}\,\,\frac{s^{(2)}_{\rm SG}(\theta)}
{\cosh\frac{(i\pi-\theta)\nu}{2}}\,,\label{sig2SG}\\
\sigma_3(\theta)&=\cosh\frac{i\pi\nu}{2}
\cosh\frac{\theta\nu}{2}\,\,\frac{s^{(2)}_{\rm SG}(\theta)}
{\cosh\frac{(i\pi-\theta)\nu}{2}}\,,\nonumber
\end{alignat}
where we have introduced the parametrization 
\be
\nu=\frac{8\pi}{\beta^2}-1.
\label{gammapar}
\end{equation}
The \lq isospin 2' phase shift for the SG model is 
\be
s^{(2)}_{\rm SG}(\theta)=-\exp\left\{2i\int_0^\infty\frac{d\omega}{\omega}
\sin\omega\theta\,\tilde k(\omega)\right\}
\label{iso2SG}
\end{equation}
with
\be
\tilde k(\omega)=\frac{\sinh\frac{\pi\omega(1-\nu)}{2\nu}}
{2\cosh\frac{\pi\omega}{2}\sinh\frac{\pi\omega}{2\nu}}.
\label{tildek}
\end{equation}
Note that in the $\beta^2\to8\pi$ ($\nu\to0$) limit the SG S-matrix
coincides with the $n\to2$ limit of the $O(n)$ S-matrix, in particular
$\lim_{\nu\to0}\tilde k(\omega)=\lim_{n\to2}\tilde K_n(\omega)$.

The identification of the XY model with the $\nu\to0$ limit of the
SG model is surprising since in this limit the bootstrap
S-matrix (\ref{sig2SG}) becomes SU(2) symmetric, coinciding with
the S-matrix of the $SU(2)$ chiral Gross-Neveu (CGN) model \cite{pew}.
It is not obvious where this enlarged symmetry comes from. 
The existence of a nontrivial XY model is even more surprising in the
light of the fact that the beta-function of the coupling $g^2$
in (\ref{LagOn}) vanishes for $n=2$ and by making the substitution
$S^1=\cos\varphi\,,\,S^2=\sin\varphi$ the Lagrangean (\ref{LagOn}) 
naively becomes free.

Kosterlitz and Thouless \cite{KT} argued that the fact that $\varphi$
is a periodic (angular) variable plays an important role and therefore
the model has nontrivial dynamics. They have shown that topologically
nontrivial objects, vortices, are present in typical spin configurations
and their interaction makes the theory nontrivial. 

The standard lattice action of the XY model is
\be
S_{\rm XY}=K\sum_{x,\mu}\big[1-
\cos\big(\varphi(x)-\varphi(x+\hat\mu)\big)\big].
\label{standardS}
\end{equation}
We denoted by $K$ the inverse of the XY model coupling to avoid confusion
with the SG coupling $\beta$. Assuming universality, not only the cosine
function but any other $2\pi$-periodic function $W(\varphi)$ which has a local 
minimum at $\varphi=0$ defines a possible XY model lattice action.
The Villain model action \cite{Villain} is chartacterized by
\be
W(\varphi)=-\frac{1}{K_V}\ln\Big[\sum_m\exp\big\{-\frac{K_V}{2}
(\varphi-2\pi m)^2\big\}\Big].
\label{VillainS}
\end{equation}

Kosterlitz and Thouless showed that typical spin configurations
can be represented as a mixture of smooth, topologically trivial
configurations (spin waves) and a gas of vortices (of integer topological
charge). The KT vortices are not interacting with the spin waves, but 
there is a logarithmic interaction potential between the vortices which
are therefore identical to a two-dimensional Coulomb gas. 
This spin wave $+$ Coulomb gas (SWCG) picture is only approximate 
if we start from the standard action (\ref{standardS}) but it is an exact 
duality transformation \cite{Jose} for the Villain action corresponding to
(\ref{VillainS}).
That the XY model with standard action is in the same universality class
as the Villain model was demonstrated using Monte Carlo renormalization
group techniques \cite{Hasenbusch}. On the other hand, it has been
shown rigorously \cite{FrohSpen} that the Coulomb gas has a phase transition
point at some finite critical coupling $K_c$. KT interpreted this phase
transition as one of vortex condensation and by a (heuristic) energy-entropy
consideration showed that in the vicinity of $K_c$ vortices of topological
charge $\pm1$ only are important, higher vortices can be neglected.
It is easy to see that this system (SWCG with unit charge vortices only) 
is exactly equivalent to the SG model. In ref. \cite{Amit} it was shown that
the extremal SG fixed point $\beta^*=\sqrt{8\pi}\,,\alpha^*=0$ is 
appropriate to discribe the KT phase transition. The renormalizabilty
of the SG model around this point was explicitly demonstrated up to
two-loop order in a simultaneous perturbative expansion in $\alpha$
and $\delta=\frac{\beta^2-8\pi}{8\pi}$.

Finally, there is a further transformation that explains the dynamical
$SU(2)$ symmetry of the XY model. The SG model can be exactly mapped 
\cite{Banksetal} to a fermionic model formulated in terms of a two-component
Dirac fermion $\psi$. The transformation is similar to the well-known one
that relates the SG model to the massive Thirring model \cite{Coleman}.
Here the fermionic model is a deformation of the chiral Gross-Neveu model 
with four-fermion interaction: 
\be
{\cal L}_F=i(\overline{\psi}\gamma_\mu\partial_\mu\psi)-
g_0(J_\mu^1)^2-g_0(J_\mu^2)^2-(g_0+f_0)(J_\mu^3)^2\,,
\label{LagF}
\end{equation}
where
\be
J_\mu^a=\frac{1}{2}\,\overline{\psi}\gamma_\mu\sigma^a\psi
\label{SU2curr}
\end{equation}
is the fermionic $SU(2)$ current. The relation between the SG couplings
$\delta\,,\alpha$ and the fermion couplings $g_0\,,f_0$ is
\be
\alpha=\frac{8}{\pi}g_0+\cdots,\qquad\qquad
\delta=-\frac{1}{\pi}(g_0+f_0)+\cdots,
\label{g0f0}
\end{equation}
where the dots indicate that the relations (\ref{g0f0}) receive higher
order corrections in perturbation theory. In the fermionic formulation
the KT fixed point is the Gaussian one and for vanishing deformation
parameter, $f_0=0$, the model is manifestly $SU(2)$ symmetric. The
corresponding relation in the SG language is $\alpha+8\delta=0$ at
lowest order. 

To summarize, the XY model in the vicinity of the Kosterlitz-Thouless 
transition point is believed to be described by the SG model with
extremal coupling $\beta=\sqrt{8\pi}$. This is further equivalent to
the two-component chiral Gross-Neveu model around its Gaussian point.
We will use the SG language throughout this paper.

\section{The SG description of the $O(2)$ model}

In this section we review the SG description of the KT theory 
closely following the approach of Amit et al. \cite{Amit}. 
Without loss of generality we can adopt the somewhat
unusual regularization scheme of the authors, 
since, as we will see, all important results are universal, 
i.e. independent of the regularization scheme. Nevertheless, it would 
be interesting to repeat all the calculations below using some of the
more customary regularizations like the lattice or dimensional regularization.

Our starting point is the Euklidean Lagrangian \cite{Amit}
\begin{equation}
{\cal L}=\frac12\partial_\mu\phi\partial_\mu\phi
+\frac{m_0^2}{2}\phi^2+\frac{\alpha_0}{\beta_0^2a^2}\left[
1-\cos(\beta_0\phi)\right],
\label{Lag}
\end{equation}
where $m_0$ is an IR regulator mass and $a$ is the UV
cutoff (of dimension length). We have denoted the dimensionless
SG couplings by $\beta_0$ and $\alpha_0$ to emphasize that they are
bare (unrenormalized). UV regularized correlation functions are 
calculated by using
\begin{equation}
G_0(x)=\frac{1}{2\pi}\,K_0\left(m_0\sqrt{x^2+a^2}\right)
\label{G0}
\end{equation}
where $K_0$ is the modified Bessel function, as the 
$\phi$ propagator. Our strategy is slightly different from
\cite{Amit}, who really considered the renormalization of
the massive SG model (\ref{Lag}) of mass $m_0$. We treat
$m_0$ as an IR regulator mass and consider IR stable
physical quantities for which we can take the limit 
${m_0\to0}$ already at the UV regularized level
(before UV renormalization). All renormalization constants are,
for example, IR stable and independent of $m_0$.

The SG coupling $\beta_0$ is close to its special value $\sqrt{8\pi}$
and a simultaneous perturbative expansion in the mass 
parameter $\alpha_0$ and the deviation $\delta_0$ is defined, where
\be
\beta_0^2=8\pi(1+\delta_0)
\end{equation}
and the parameters are renormalized according to
\ba
\alpha_0&=&Z_\alpha\alpha\,,\\
\beta_0^2&=&Z_\phi^{-1}\beta^2\,.
\ea
Here the Z-factors are functions of the renormalized couplings
$\alpha$ and $\delta$ and the combination
\be
l=\ln\mu a\,,
\end{equation}
where $\mu$ is an arbitrary mass parameter (basically the normalization
point). Similarly, a renormalization constant $Z$ is necessary to make
${\cal G}$, the spin-spin 2-point function finite:
\be
{\cal G}_R=Z{\cal G}\,.
\end{equation}

For vanishing mass parameter $\alpha$ the Lagrangian (\ref{Lag}) is trivial
and $Z_\phi=1$ since there is no need to renormalize the SG coupling. The
spin-spin correlation function (which is an exponential of
the basic field $\phi$) gets renormalized even in this point, but in this
case its renormalization constant is simply
\be
Z=(\mu a)^{-{1\over4(1+\delta)}}\,.
\end{equation}
In addition, there is a symmetry $\alpha\leftrightarrow-\alpha$ 
(which corresponds to a $\pi\over\beta_0$ shift in the basic field). 

Taking into account the above constraints, the perturbative expansion
of the Z-factors must be of the form
\ba
Z_\phi&=&1+f_1\alpha^2l+\alpha^2\delta(\bar f_2l^2+f_2l)+\cdots\,,\\
Z_\alpha&=&1+g_1\delta l+\alpha^2(\bar g_2l^2+g_2l)+\delta^2(\bar g_3l^2+
g_3l)+\cdots\,,\\
Z&=&e^{-{l\over4(1+\delta)}}\big\{1+\alpha^2(\bar h_1l^2+h_1l)+
\cdots\big\}\,.
\ea
Amit et al. \cite{Amit} found the following results.
\ba
f_1={1\over32}\,,\qquad\qquad\bar f_2&=&-{1\over16}\,,\qquad\qquad\ \ 
f_2=-{3\over32}\,,\nonumber\\
g_1=-2\,,\qquad\qquad\bar g_2&=&\phantom{-}{1\over32}\,,\qquad\qquad\ \  
g_2=-{5\over64}\,,\label{Am2}\\
\bar g_3&=&\phantom{+}\ 2\,,\qquad\qquad\ \ \  g_3=
\phantom{-}\ 0\,.\nonumber
\ea
Furthermore $\bar h_1=-1/256$, but the number $h_1$ is not known at present.
The above two-loop beta-function coefficients were calculated also by
other methods. The original calculation has recently been reconsidered and
the results (\ref{Am2}) have been confirmed \cite{BH}.

The spin-spin correlation function satisfies the equation
\be
{\cal D}{\cal G}=\gamma{\cal G}\,,
\label{RGE}
\end{equation}
where ${\cal D}$ is the renormalization group (RG) operator 
\be
{\cal D}=-a{\partial\over\partial a}+D=
-a{\partial\over\partial a}
+\beta_\alpha(\alpha_0,\delta_0){\partial\over\partial\alpha_0}
+\beta_\delta(\alpha_0,\delta_0){\partial\over\partial\delta_0}
\end{equation}
and the $\beta$ and $\gamma$-functions are given by
\ba
\beta_\alpha(\alpha_0,\delta_0)&=&-g_1\alpha_0\delta_0-g_2\alpha_0^3-
g_3\alpha_0\delta_0^2+\cdots\,,\label{beta1}\\
\beta_\delta(\alpha_0,\delta_0)&=&f_1\alpha_0^2+(f_1+f_2)\alpha_0^2\delta_0
+\cdots\,,\label{beta2}\\
\gamma(\alpha_0,\delta_0)&=&-{1\over4}+{1\over4}\delta_0-{1\over4}\delta_0^2+
h_1\alpha_0^2+\cdots\,.\label{gamma}
\ea

Now, as is well known, not all $\beta$-function coefficients are universal.
For example, under a redefinition
\ba
\tilde\alpha_0&=&\alpha_0+L\alpha_0\delta_0+\cdots\,,\label{Redef1}\\
\tilde\delta_0&=&\delta_0+K\alpha_0^2+\cdots\,,\label{Redef2}
\ea
the coefficients change as
\ba
\tilde g_1&=&g_1\,,\phantom{-Kg_1-Lf_1,}\qquad\qquad\tilde f_1=f_1\,,
\nonumber\\
\tilde g_2&=&g_2-Kg_1-Lf_1\,,\qquad\qquad\tilde f_2=f_2
-2Kg_1-2Lf_1\,,\label{redef2}\\
\tilde g_3&=&g_3\,.\nonumber
\ea
((\ref{Redef1}-\ref{Redef2}) is the most 
general perturbative redefinition respecting 
the $\alpha_0\leftrightarrow-\alpha_0$ symmetry together with the requirement 
that for $\alpha_0=0$ $\delta_0$ is not redefined.)
From (\ref{redef2}) we see that
in addition to the one-loop coefficients $g_1$ and $f_1$ there exist also
two-loop invariants. They are $g_3$ and the combination
\be
J=2g_2-f_2\,.
\label{J}
\end{equation}

Two important physical quantities are the correlation length
\be
\xi={1\over Ma}=e^{\Psi(\alpha_0,\delta_0)}\,,
\label{xi}
\end{equation}
where $M$ is the mass of the physical particle
and the dimensionless susceptibility
\be
\chi={1\over a^2}\int d^2 x\,{\cal G}(x)=e^{\Phi(\alpha_0,\delta_0)}\,.
\label{chi}
\end{equation}
From (\ref{RGE}) it follows that the exponents satisfy
\ba
D\Psi&=&1\,,\label{Psi}\\
D\Phi&=&2+\gamma\,.\label{Phi}
\ea
 
It is useful to introduce the RG invariant quantity $Q(\alpha_0,\delta_0)$
which satisfies $DQ=0$. Introducing the inverse function $k(\delta_0,Q)$
that satisfies
\be
\alpha_0=k(\delta_0,Q(\alpha_0,\delta_0))\,,
\end{equation}
we can define new $\beta$- and $\gamma$-functions:
\ba
b(\delta_0,Q)&=&\beta_\delta(k(\delta_0,Q),\delta_0)\,,\label{b}\\ 
\Gamma(\delta_0,Q)&=&\gamma_\delta(k(\delta_0,Q),\delta_0)\,.\label{Gamma}
\ea
The advantage of using the variables $\delta_0$ and $Q$ is that the RG
invariant $Q$ can be treated in many respects as if it were a numerical
constant and $\delta_0$ were a single coupling constant. For example,
if we write
\be
\Psi(\alpha_0,\delta_0)=\Psi_1(\delta_0,Q(\alpha_0,\delta_0))
\end{equation}
and
\be
\Phi(\alpha_0,\delta_0)=\Phi_1(\delta_0,Q(\alpha_0,\delta_0))
\end{equation}
then the functions $\Psi_1$ and $\Phi_1$ can be determined from
\be
{\partial\over\partial\delta_0}\Psi_1(\delta_0,Q)={1\over b(\delta_0,Q)}
\label{Psi1diff}
\end{equation}
and
\be
{\partial\over\partial\delta_0}\Phi_1(\delta_0,Q)=
{2+\Gamma(\delta_0,Q)\over b(\delta_0,Q)}
\label{Phi1diff}
\end{equation}
respectively.

Using (\ref{beta1}) and (\ref{beta2}) $Q$ can be determined. 
\be
Q(\alpha_0,\delta_0)=\frac{1}{32}\alpha_0^2-2\delta_0^2+
2g_2\alpha_0^2\delta_0+F_2\delta_0^3+\cdots
\label{Q}
\end{equation}
and using this in (\ref{b}) we find
\be
b(\delta_0,Q)=Q+2\delta_0^2+AQ\delta_0+B\delta_0^3+\cdots\,,
\label{bb}
\end{equation}
while the $\Gamma$ function to this order is
\be
\Gamma(\delta_0,Q)=-{1\over4}+{1\over4}\delta_0+\cdots\,.
\end{equation}
Here
\be
A=1-\frac{J}{f_1}\,,\qquad\qquad
B=-\frac{2}{3}g_3+\frac{2}{3}A\,,\qquad\qquad
F_2=\frac{2}{3}g_3+\frac{4}{3}A
\label{ABF2}
\end{equation}
or numerically
\be
A=3\,,\qquad\qquad B=2\,,\qquad\qquad F_2=4\,,
\label{ABF2num}
\end{equation}
but we will keep these constants in the following to explicitly
demonstrate that all our results are universal.

We know that a one-parameter renormalizable subspace 
in the $\delta_0$--$\alpha_0$ plane is equivalent to the SU(2) 
chiral Gross-Neveu model. (This is most evident in the fermionic formulation.)
This subspace must correspond to the $Q=0$ RG trajectory
because we know that it goes through the point 
$\alpha_0=\delta_0=0$. Moreover, it must be the $\delta_0<0$ branch
of the $Q=0$ trajectory, since it is the one that is asymptotically
free in perturbation theory.

\begin{figure}[htb]
\leavevmode
\epsfxsize=170mm
\hspace{-1.0cm}
\epsfbox{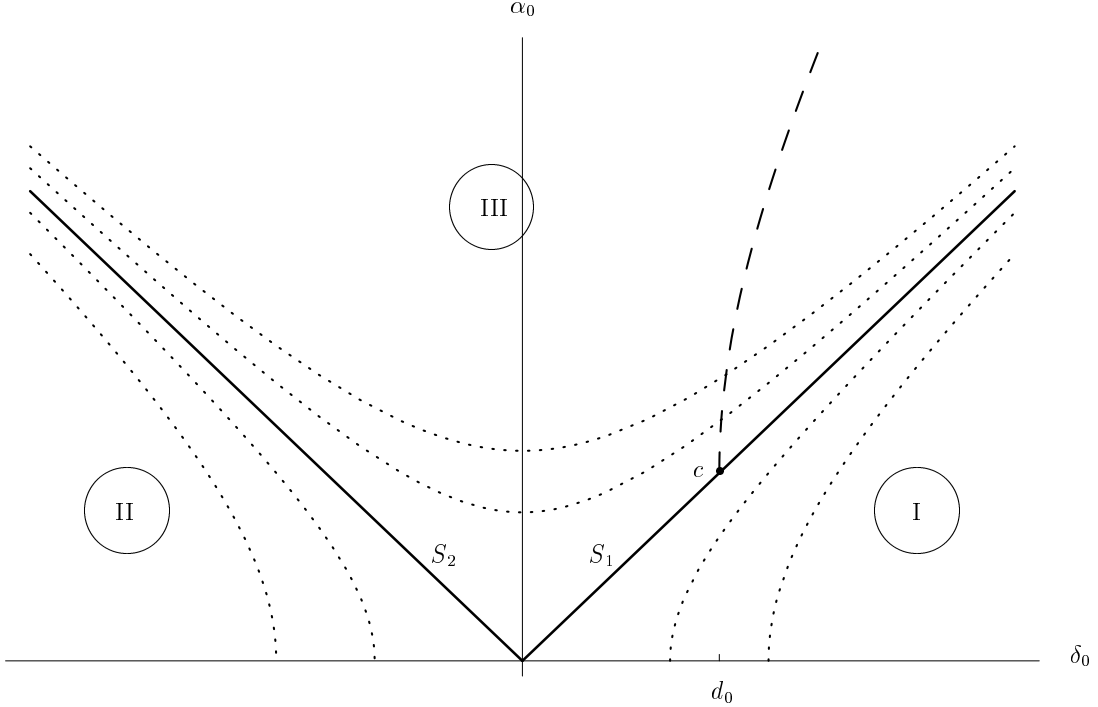}

\vspace{0.1cm}

\caption{\footnotesize
Phase diagram of the KT theory. The entire Region I is
critical. The model is massive in Region III ($Q>0$) and
Region II ($Q<0$). The dotted lines are RG trajectories.
The full lines $S_1$ and $S_2$ correspond to the critical
surface of the XY model and the (bare) CGN model respectively.
The dashed curve is the XY model hitting the critical surface
at $c$.
} 
\label{extra1}
\end{figure}

Following \cite{Amit} the phase diagram of our model is represented
in Figure 1. 
The CGN model corresponds to the separatrix $S_2$ on this plot,
which will be referred to as PD for short. Region III 
corresponds to $Q>0$, whereas Regions I and II correspond to 
$Q<0,\delta_0>0$ and $Q<0,\delta_0<0$, respectively. 

In the neighbourhood of $S_2$, i.e. close to the CGN case
we have
\be
b(\delta_0,Q)=b_0(\delta_0)+Qb_1(\delta_0)+\cdots\,,
\end{equation}
where
\ba
b_0(\delta_0)&=&2\delta_0^2+B\delta_0^3+\cdots\,,\label{b0}\\
b_1(\delta_0)&=&1+A\delta_0+\cdots\,.\label{b1}
\ea
It is a crucial observation that near the CGN line the correlation
length exponent $\Psi_1$ is a smooth analytic function of $Q$:
\be
\Psi_1(\delta_0,Q)=H_0(\delta_0)+QH_1(\delta_0)+\cdots\,,
\label{Psi1CGN}
\end{equation}
where
\ba
H_0(\delta_0)&=&-{1\over2\delta_0}-\frac{B}{4}\ln\vert2\delta_0\vert+
\Psi_0+\cdots\,,\\
H_1(\delta_0)&=&{1\over12\delta_0^3}+\frac{A-B}{8\delta_0^2}+\cdots\,.
\ea
We will calculate the value of the nonperturbative constant $\Psi_0$ 
in Section 5.

Now we can integrate (\ref{Psi1diff}) using the perturbative expansion
(\ref{bb}) and for $Q>0$ we get 
\ba
\Psi_1(\delta_0,Q)&=&{1\over\sqrt{2Q}}\Bigg({\pi\over2}+{\rm arctg}
{\sqrt{2}\delta_0\over\sqrt{Q}}\Bigg)-\frac{B}{8}\ln2(Q+2\delta_0^2)
\nonumber\\
&+&\frac{2A-B}{8}{Q\over Q+2\delta_0^2}+\Psi_0+\cdots\,.\label{pertI}
\ea
Here the dots stand for higher order terms in the perturbative expansion
(these are, in principle, calculable) and also for an unknown 
(nonperturbative) function of $Q$ only. For small $Q$ (\ref{pertI}) 
becomes
\be
\Psi_1(\delta_0,Q)={\pi\over\sqrt{2Q}}\,\Theta(\delta_0)+
H_0(\delta_0)+QH_1(\delta_0)+\cdots\,,
\label{smallQ}
\end{equation}
where the dots stand for terms higher order in Q. They come from the
higher perturbative terms of (\ref{pertI}) and also from the 
nonperturbative function mentioned above. The point is that there are
no terms, singular in $Q$, coming from any of these two sources. This
is obvious for the perturbative terms, but must also be true for the
nonperturbative contributions, since otherwise $\Psi_1$ would be
singular on the $S_2$ line. Requiring it to be nonsingular on $S_2$,
we force $\Psi_1$ to diverge on $S_1$, which is therefore part of the
critical surface of the phase diagram PD. An other line, on which we 
know the correlation length must diverge is the $\delta_0$
axis $\alpha_0=0$, because this corresponds to a free, massless model.

For $Q<0$, it is convenient to parametrize $Q$ in terms of the 
$\delta_0$ value at which the RG trajectory intersects this axis.
In other words, we have to express $Q$ in terms of $d$ that solves
\be
b(d,Q)=0\,.
\label{d}
\end{equation}
(Note that in this parametrization $\vert\delta_0\vert\geq\vert d\vert$, 
because of $\alpha_0^2\geq0$.)
The perturbative solution of (\ref{d}) is
\be
Q=-2d^2+(2A-B)d^3+\cdots\,.
\label{Qsol}
\end{equation}
Using this parametrization the perturbative solution for $Q<0$ is
\ba
\Psi_1(\delta_0,Q)&=&-{1\over4d}\ln\Big({\delta_0+d\over\delta_0-d}
\Big)\Big[1+\frac{2A-3B}{4}d\Big]
+\frac{2A-B}{8}{d\over\delta_0+d}\nonumber\\
&-&\frac{B}{4}\ln2\vert\delta_0+d\vert+y(d)+\cdots\,,\label{pertII}
\ea
where
\ba
{\rm for\ }d<0\qquad y(d)&=&\Psi_0+\cdots\,,\label{dneg}\\
{\rm for\ }d>0\qquad y(d)&=&\infty\,.\label{dpos}
\ea
(\ref{dneg}) is obtained by matching (\ref{pertII}) to
(\ref{Psi1CGN}) for small $Q$, while (\ref{dpos}) is formally true
since this is the only way to achieve that the correlation length
diverges on the positive half of the $\delta_0$ axis. 
(Without the infinite constant $\xi$ actually {\it vanishes}
there.)

Now we can discuss the phase diagram of our model. The entire Region I
is critical. The massive phase is Regions II and III and the critical
surface bordering them is $S_1$ plus the negative part of the $\delta_0$
axis. They are smoothly connected across $S_2$, which is the (bare)
CGN model. The $O(2)$ NLS model corresponds to the dashed curve of PD.
In a MC experiment we are approaching the critical point $c$ from the
massive phase (Region III). We will denote the $\delta_0$ coordinate of
$c$ by $d_0$. Because the RG trajectories are running
basically parallel to $S_1$, it is physically irrelevant at which point
the critical surface is reached and therefore the parameter $d_0$ is
irrelevant. The continuum model will be the same for all points on
$S_1$, including the origin. But the origin is the point, where
(coming along $S_2$) the continuum CGN model is defined! So our
continuum theory is inevitably identical to the (massive part of the)
$SU(2)$-invariant CGN model.

If we start from somewhere in the middle of Region II, we can define
a massive continuum limit by approaching the negative half of the
$\delta_0$ axis. The intercept $d$ is then relevant. The continuum
theory is the SG model with 
\be
\beta^2=8\pi(1+d).
\label{betad}
\end{equation}

Returning to the $O(2)$ model, the dashed trajectory can be parametrized
as
\ba
\delta_0&=&d_0+d_1\tau+\cdots\,,\\
\alpha_0&=&k(d_0,0)+\alpha_1\tau+\cdots\,,
\ea
where 
\be
\tau=K_c-K
\label{tau}
\end{equation}
is the reduced coupling and we have assumed that physical
quantities are analytic in $K$. ($K_c=1.1197(5)$ 
\cite{Hasenbusch}.) Then also $Q$ is analytic in $\tau$:
\be
Q\sim\tau\,.
\end{equation}
From (\ref{smallQ}) we see that along the $O(2)$ curve
\be
\ln\xi={\pi\over\sqrt{2Q}}+H_0(d_0)+\cdots\,,
\label{logxi}
\end{equation}
where the dots stand for terms analytic in $Q$ (and vanishing for $Q=0$).
In other words, for the $O(2)$ model \cite{KT}
\be
\xi=C\exp\Big\{\frac{b}{\sqrt{\tau}}\Big\}\big(1+
{\cal O}(\sqrt{\tau})\big),
\label{xiKT}
\end{equation}
where the constants $C$ and $b$ are not universal. This is the famous KT
formula showing that the phase transition is of infinite order in the
reduced temperature.

It is more important for us that (\ref{logxi}) can be rewritten as
\be
Q={\pi^2\over2(\ln\xi +u)^2}+{\cal O}\big((\ln\xi)^{-5}\big),
\label{Qdef}
\end{equation}
where
\be
u=-H_0(d_0)\,,
\label{u}
\end{equation}
which is given (if $d_0$ is sufficiently small) perturbatively by
\be
u\approx{1\over2 d_0}+\frac{B}{4}\ln(2d_0)-\Psi_0\,.
\end{equation} 
Note that the leading $1/(\log\xi)^2$
term in (\ref{Qdef}) is universal and only the subleading terms 
(depending on the value of the parameter $u$) are model-dependent.

The susceptibility exponent $\Phi_1$ can be studied similarly. It can
be written as
\be
\Phi_1={7\over4}\,\Psi_1+\tilde\Phi_1={7\over4}\,\Psi_1+
{1\over16}\ln(2\delta_0^2+Q)+\cdots\,,
\end{equation}
where $\tilde\Phi_1$ satisfies
\be
{\partial\tilde\Phi_1\over\partial\delta_0}={\Gamma(\delta_0,Q)+{1\over4}
\over b(\delta_0,Q)}\,.
\end{equation}

Now the crucial observation is again that, for small $Q$,
\be
\Phi_1={7\over4}\,\Psi_1+c_1+c_2Q+\cdots\,,
\label{Phi1smallQ}
\end{equation}
because the (calculable) perturbative terms are analytic in $Q$, while 
the (not calculable) purely $Q$-dependent terms in $\tilde\Phi_1$ must
also be analytic in $Q$ otherwise these latter singularities would
also turn up for the CGN line $S_2$, where they must not.

From (\ref{Phi1smallQ}) we have
\be
\chi=\xi^{{7\over4}}\,e^{c_1}\Big(1+c_2Q+\cdots\Big)\,,
\label{chixi}
\end{equation}
i.e. there are no (multiplicative) log corrections in the scaling
relation for the susceptibility. The possibility of such
multiplicative logarithmic corrections is discussed in \cite{JankeKI}.

\section{Determination of the lattice artifacts}

Recall that the RG invariant $Q$ has
a completely different meaning for Region III (which contains the
massive phase of the XY model) and for Region II (where the usual
massive SG model with $\beta^2<8\pi$ can be defined). Indeed, in
the positive $\delta_0$  part of Region III, 
close to $S_1$, the (positive) parameter $Q$ merely measures the 
distance from the XY
critical surface on which it vanishes, whereas in Region II $Q$ is
a relevant (negative) parameter related to the SG coupling $\beta$ 
by (\ref{d}) and (\ref{betad}). 
Our main assumption is that in spite of this difference
physical quantities are smoothly depending on $Q$ in the vicinity of the 
separatrix $S_2$ connecting the two regions. More precisely, we will
assume that close to $S_2$ any physical quantity $U$ has the form
\be
U(Q)=U_0+U_1Q+{\cal O}(Q^2).
\label{Qexp}
\end{equation}
Here $U_0=U(0)$ is its value for the CGN model (and thus also in the
continuum limit of the XY model). The first correction coefficient $U_1$
can be calculated from the SG model as follows. Using the identification
(\ref{d}) and its perturbative solution (\ref{Qsol}) together with
(\ref{betad}) and (\ref{gammapar}) we have
\be
Q=-2\nu^2 +{\cal O}(\nu^4).
\label{Qgamma}
\end{equation}
This means that if in the SG model, close to the CGN point $\nu=0$,
we have
\be
U(\nu)=u_0+u_1\nu^2+{\cal O}(\nu^4),
\label{gammaexp}
\end{equation}
then
\be
U_0=u_0\qquad\qquad{\rm and}\qquad\qquad U_1=-u_1/2.
\label{u0u1}
\end{equation}
Translated to the language of lattice artifacts by (\ref{Qdef}) we thus
have
\be
U(\xi)=u_0-\frac{u_1\pi^2}{4(\ln\xi+u)^2}+{\cal O}\big((\ln\xi)^{-4}\big).
\label{xiexp}
\end{equation}
This means that lattice artifacts typically go away very slowly, only as
$1/(\log\xi)^2$. On the other hand the leading artifacts are universal and
calculable.

We apply this method first to the scattering phase shifts. Recall 
the SG model S-matrix (\ref{OnSmatrix}) with (\ref{sig2SG}).
The three distinct S-matrix eigenvalues are
\ba
s^{(0)}_{\rm SG}(\theta)=2\sigma_1(\theta)+\sigma_2(\theta)+\sigma_3(\theta)
&=&\phantom{-}\frac
{\sinh\frac{\nu}{2}(i\pi+\theta)}
{\sinh\frac{\nu}{2}(i\pi-\theta)}s^{(2)}_{\rm SG}(\theta),\label{s0}\\
s^{(1)}_{\rm SG}(\theta)=\sigma_2(\theta)-\sigma_3(\theta)
&=&-\frac
{\cosh\frac{\nu}{2}(i\pi+\theta)}
{\cosh\frac{\nu}{2}(i\pi-\theta)}s^{(2)}_{\rm SG}(\theta)\label{s1}
\ea
and
\be
s^{(2)}_{\rm SG}(\theta)=\sigma_2(\theta)+\sigma_3(\theta)
=-\exp\Big\{2i\int_0^\infty\frac{d\omega}{\omega}
\sin(\theta\omega)\,\tilde k(\omega)\Big\}
\label{s2}
\end{equation}
where $\tilde k(\omega)$ is given by (\ref{tildek}).

We now write
\be
s^{(i)}_{\rm SG}(\theta)=\exp\big\{2i\delta_0^{(i)}(\theta)+2i
\nu^2\delta_1^{(i)}(\theta)+{\cal O}(\nu^4)\big\}
\label{phaseshifts}
\end{equation}
for $i=0,1$ and $2$. Here $\delta^{(i)}_0$ are the CGN phase shifts
which, as remarked before, coincide with the $n\to2$ limit of the
$O(n)$ phase shifts. 
The first correction coefficients can be obtained by a simple
calculation. The result is 
\be
\delta^{(0)}_1(\theta)=0,\qquad
\delta^{(1)}_1(\theta)=\frac{\pi\theta}{6},\qquad
\delta^{(2)}_1(\theta)=-\frac{\pi\theta}{12}.
\label{1correction}
\end{equation}
This can be used to obtain the leading lattice artifacts in the
XY model by the relation (\ref{xiexp}).

\section{Current-current 2-point function and free energy}

Consider the 2-point function of the Noether current 
\be
J_\mu=i{\beta_0\over2\pi}\epsilon_{\mu\nu}\partial_\nu\phi\,.
\label{defJ}
\end{equation} 
Its Fourier transform $I(p)$ is defined by
\be
\langle J_\mu(x)J_\nu(y)\rangle=\int {d^2p\over(2\pi)^2}\,
e^{-ip(x-y)}{p_\mu p_\nu-p^2\delta_{\mu\nu}\over p^2}\,I(p)\,.
\end{equation}
It is easy to calculate $I(p)$ to second order:
\be
I(p)={2\over\pi}\Bigg\{1+\delta_0+\alpha_0^2\Big[\Omega(p,a)+\cdots
\Big]\Bigg\}\,,
\label{curr}
\end{equation}
where
\be
\Omega(p,a)=f_1(\ln pa +\Omega_0)
\label{Omega}
\end{equation}
and 
\be
\Omega_0=C-{1\over2}-\ln2\,,
\label{Omega0}
\end{equation}
$C$ being Euler's constant. 

Standard RG considerations give
\be
I(p,\delta_0,Q,a)=I(p_0,\bar\lambda,Q,a)=\tilde I(p/M,Q)\,,
\label{RGI}
\end{equation}
where the running coupling is the solution of
\be
\Psi_1(\bar\lambda,Q)=\ln{p\over M}-\ln p_0a\,.
\label{lambdabar}
\end{equation}
 
Let us consider the $Q<0$, $\delta_0<0$ case (Region II) first.
For $p\rightarrow\infty$ also $\Psi_1\rightarrow\infty$ and therefore
$\bar\lambda\rightarrow d$, where $d$ is defined by
\be
b(d,Q)=0\qquad\qquad{\rm or\ equivalently}
\qquad\qquad\alpha_0=k(d,Q)=0\,.
\end{equation}
This gives
\be
\tilde I(\infty,Q)={2\over\pi}(1+d)\,,
\label{Iinfty}
\end{equation}
where $d$ parametrizes $Q$ according to (\ref{Qsol}). 
(\ref{Iinfty}) is consistent with the identification (\ref{betad}).

Next we study the case of small $Q$, because this is relevant if
we are interested in the approach to the continuum limit along
the $O(2)$ curve. We assume that $\tilde I$ is analytic in $Q$:
\be
\tilde I(z,Q)=\tilde I_0(z)+Q\tilde I_1(z)+\cdots\,.
\label{Iexpand}
\end{equation}
It is a standard exercise to obtain the asymptotic
expansion of the coefficients $\tilde I_0$ and $\tilde I_1$ in perturbation
theory. One gets
\ba
\tilde I_0(p/M)&=&{2\over\pi}\Bigg\{1-\lambda+2\kappa\lambda^2+\cdots\Bigg\}
\,,\label{I0}\\ 
\tilde I_1(p/M)&=&{2\over3\pi}\Bigg\{{1\over2\lambda}+\kappa
+\frac{2B-3A}{4}+\cdots\Bigg\}\,,\label{I1}
\ea
where
\be
\kappa=\Omega_0-\Psi_0
\label{kappa}
\end{equation}
and the coupling $\lambda$ is the solution of
\be
{1\over2\lambda}-\frac{B}{4}\ln(2\lambda)=\ln{p\over M}\,.
\end{equation}
The asymptotic expansions (\ref{I0}) and (\ref{I1}) are valid for
$\lambda\ll1$, i.e. for $p\rightarrow\infty$ but also $Q$ must
satisfy
\be
Q\ll6\lambda^2
\label{Qsmaller}
\end{equation}
so that the expansion (\ref{Iexpand}) makes sense.

It is by now standard how the nonperturbative constant $\kappa$
can be calculated. For this it is necessary to consider the free
energy in an external field and we now turn to this calculation.
We follow here \cite{Zamolodchikov} and start from the modified
Lagrangian
\be
{\cal L}_h={1\over2}(\partial_\mu\phi)^2-{\alpha_0\over\beta_0^2 a^2}
\cos(\beta_0\phi)+{\beta_0 h\over2\pi}\partial_1\phi\,,
\label{Lagh}
\end{equation}
which corresponds to adding a term $ihJ_2$ to the Lagrangian density.
The modified ground state energy must be of the form
\be
{\cal E}(h)=-{h^2\over\pi}\,{\cal F}(h)\,,
\end{equation}
where ${\cal F}(h)$ is dimensionless.

In perturbation theory we get
\be
{\cal F}(h)=1+\delta_0+\alpha_0^2\Big[\tilde\Omega(h,a)+\cdots
\Big]\,,
\label{calF}
\end{equation}
where
\be
\tilde\Omega(h,a)=f_1(\ln ha +\tilde\Omega_0)
\label{TOmega}
\end{equation}
and 
\be
\tilde\Omega_0=C-{1\over2}+\ln2\,.
\label{TOmega0}
\end{equation}
For $Q=0$ therefore ${\cal F}(h)$ has the following asymptotic expansion
\be
{\cal F}(h)=1-{1\over2\ln{h\over M}}-\frac{B}{8}
{\ln\ln {h\over M}\over\ln^2{h\over M}}+
{\kappa+\ln4\over2\ln^2{h\over M}}+
{\cal O}\Bigg({\Big(\ln\ln {h\over M}\Big)^2\over\ln^3{h\over M}}
\Bigg)\,.
\label{pertF}
\end{equation}
This is to be compared to the exact result \cite{Forgacs}
\be
{\cal F}(h)=1-{1\over2\ln{h\over M}}-
{\ln\ln {h\over M}\over4\ln^2{h\over M}}+
{{3\over2}\ln2-{1\over4}-{1\over2}\ln\pi
\over2\ln^2{h\over M}}+
{\cal O}\Bigg({\Big(\ln\ln {h\over M}\Big)^2\over\ln^3{h\over M}}
\Bigg)\,.
\label{exF}
\end{equation}
It is a nontrivial check on the overall consistency of our considerations
that using (\ref{ABF2num}) for the numerical value of our 
universal constant $B$ the third term in (\ref{pertF}) matches the
corresponding one in (\ref{exF}).
Comparing (\ref{pertF}) to (\ref{exF}) we also get
\be
\kappa=-{1\over4}-\ln\sqrt{2\pi}\,.
\label{kappaEX}
\end{equation}
Using the exact value (\ref{kappaEX}) we obtain the asymptotic
expansion of the first correction in $Q$:
\be
\tilde I_1(p/M)={2\over3\pi}\Bigg\{{1\over2\lambda}-{3\over2}-
\ln\sqrt{2\pi}\Bigg\}={2\over3\pi}\Bigg\{{1\over2\lambda}-2.419
\Bigg\}\,.
\label{deltaI}
\end{equation}
Note that the leading correction term (\ref{deltaI}) does not contain any
free parameter.


\vspace{1cm}
{\tt Acknowledgements}

\noindent 
This investigation was supported in part by the Hungarian National 
Science Fund OTKA (under T030099 and T029802). I thank the members of the
\lq$\sigma$ collaboration',
M. Niedermaier, F. Niedermayer, A. Patrascioiu, E. Seiler and P. Weisz,
for many thorough and interesting discussions. I also thank the 
Max-Planck-Institut f\"ur Physik for hospitality.

\end{document}